\documentclass[a4paper,11pt]{article}
\pdfoutput=1 

\usepackage{jheppub} 

\usepackage[T1]{fontenc} 

\title{\boldmath Noncommutative Cubic Scalar Field Theory and Tree Level Processes}

\author[a]{Nikhil Kalyanapuram}

\affiliation[a]{Chennai Mathematical Institute,\\Building H1, SIPCOT, Siruseri, Kelambakkam, India}

\emailAdd{nikhilkaly@cmi.ac.in}

\abstract{In this paper, we study the corrections to tree level scattering that arise due to noncommutative deformations of cubic scalar field theory through implementation of the Groenewald-Moyal(GM) product. The additional noncommutative refinement as we shall see, supplies nontrivial corrections to scattering even at the tree level. The amplitudes for $2\rightarrow2$ scattering are computed and elucidated. Finally, we also consider the possibility of a localized noncommutativity parameter and the possible consequences of introducing position dependence in spacetime noncommutativity.}

\begin{document} 
\maketitle
\flushbottom

\section{Introduction}
\label{sec:intro}

Introducing spacetime noncommutativity essentially begins with a refinement of the standard Heisenberg uncertainty principle. Instead of insisting on a nontrivial commutation relation between canonically conjugate variables, one may introduce a nontrivial commutation identity satisfied by all parameters of the theory. The most common implementation of this idea is carried out by suggesting a spacetime relation of the form,

\begin{equation}
    [x^{\mu},x^{\nu}] = 2\theta^{\mu\nu}.
\end{equation}

One way to immediately impose this commutation relation is by a modification or deformation of the multiplication operator. Standard pointwise multiplication is replaced by a new $\star$ operator which is defined by the series,

\begin{equation}
    f\star g = fg + \theta^{\mu\nu}\partial_{\mu}f\partial_{\nu}g + \theta^{\mu\nu}\theta^{\alpha\beta}\partial_{\mu}\partial_{\alpha}f \partial_{\nu}\partial_{\beta}g + ...
\end{equation}

It is clear that when the difference,

$$
x^{\mu}\star x^{\nu} - x^{\nu}\star x^{\mu}
$$

is computed, one recovers precisely the commutation relation described by $(1.1)$.

It is instructive to note here that no qualitative change has been introduced in the analytic structure of the functions involved. It is the product of two functions at a given spacetime point which now behaves differently, a consequence of the spacetime noncommutativity. 

Noting that one the product has been expanded suitably, the resulting series in a $c$ numbered function of the coordinates, we may employ the standard principles of quantum field theory, including the Feynman rules for computing scattering amplitudes. 

In the following section, we will generalise cubic scalar field theory to accommodate the new definition of the product and study the influence of the leading terms on the scattering amplitudes at tree level.

\section{Cubic Scalar Field Theory and Leading Order Corrections}

Consider the standard lagrangian for scalar field theory coupled to a cubic term,

\begin{equation}
    \mathcal{L} = -\frac{1}{2}\partial_{\mu}\phi\partial^{\mu}\phi - \frac{1}{2}m^2\phi^2 + \lambda\frac{1}{3!}\phi^3
\end{equation}

We would like to now study corrections to the lagrangian obtained by generalising the above lagrangian to the case where multiplication is given by the star product. The exact lagrangian in the new theory is given by,

\begin{equation}
    \mathcal{L} = -\frac{1}{2}\partial_{\mu}\phi\star\partial^{\mu}\phi - \frac{1}{2}m^2\phi\star\phi + \lambda\frac{1}{3!}\phi\star\phi\star\phi
\end{equation}

Now, we will study the effect that this modification will have on the Feynman rules and concordantly, on the scattering amplitudes. Since we will focus on only tree level calculations and will be avoiding loops, we will stick to the lowest order contributions.

It may immediately be noted that there are no corrections to the propagator on account of the noncommutative deformation. This is seen easily by considering the term,

\begin{equation}
    \partial^{\mu}\phi \star \partial_\mu \phi.
\end{equation}

Consider the leading noncommutative correction,

\begin{equation}
    \theta^{\alpha\beta}\partial_{\alpha}\partial_{\mu}\phi \partial_{\beta}\partial^{\mu}\phi.
\end{equation}

The Pauli-Jordan function $D(x)$ arises as a solution to the equations of motion. Successively integrating the above term by parts and bringing all the derivatives to act on only one of the field strengths allows one to argue that the terms does not contribute to the action. This may be argued by noting the antisymmetry of $\theta$ contrasted with the commutativity of the partial derivatives, implying the vanishing of the above term in the action.

Analogous arguments may immediately be made with regards to the higher order terms and the corrections to the mass term, clearly showing that no correction is made to the propagator on account of the noncommutativity. 

Now, we will study the corrections to the cubic term, which are actually nontrivial. First, we note that the star product is associative. Hence, we first evaluate the two point product,

$$
\phi\star\phi
$$

This expands out to give,

\begin{equation}
    \phi^2 + \theta^{\mu\nu}\partial_{\mu}\phi\partial_{\nu}\phi + \theta^{\mu\nu}\theta^{\alpha\beta}\partial_{\mu}\partial_{\alpha}\phi\partial_{\nu}\partial_{\beta}\phi + ...
\end{equation}

We only treat the theory upto second order in the noncommutativity parameter. Hence, only the above three terms will be counted. Let it be noted that the second term above vanishes by virtue of the antisymmetry of $\theta^{\mu\nu}$ and the symmetry of the tensor $\partial_{\mu}\phi\partial_{\nu}\phi$.

Operating $\phi \star$ on the above expansion, we now want to collect those terms which constitute the second order correction to the lagrangian. These are,

\begin{equation}
    2\theta^{\mu\nu}\theta^{\alpha\beta}\partial_{\mu}\partial_{\alpha}\phi\left(\phi\partial_{\nu}\partial_{\beta}\phi + \partial_{\beta}\phi\partial_{\nu}\phi\right)
\end{equation}
 
 and
 
\begin{equation}
    \phi\theta^{\mu\nu}\theta^{\alpha\beta}\partial_{\mu}\partial_{\alpha}\phi\partial_{\nu}\partial_{\beta}\phi.
\end{equation}

Note that the term $\theta^{\mu\nu}\partial_{\mu}\phi\partial_{\nu}(\phi^2)$ again vanishes on account of the same reasons as given earlier. Now, collecting the second order terms gives us the following second order correction to the lagrangian,

\begin{equation}
    \delta\mathcal{L}^{(2)} = \frac{\lambda}{3!}\left(3\phi\theta^{\mu\nu}\theta^{\alpha\beta}\partial_{\mu}\partial_{\alpha}\phi\partial_{\nu}\partial_{\beta}\phi + 2\theta^{\mu\nu}\theta^{\alpha\beta}\partial_{\mu}\partial_{\alpha}\phi \partial_{\beta}\phi\partial_{\nu}\phi\right)
\end{equation}

Finally, the interaction hamiltonian takes the form,

\begin{equation}
    \mathcal{H}_{I} = - \frac{\lambda}{3!}\left(\phi^3 + 3\phi\theta^{\mu\nu}\theta^{\alpha\beta}\partial_{\mu}\partial_{\alpha}\phi\partial_{\nu}\partial_{\beta}\phi + 2\theta^{\mu\nu}\theta^{\alpha\beta}\partial_{\mu}\partial_{\alpha}\phi \partial_{\beta}\phi\partial_{\nu}\phi\right)
\end{equation}

From this, we now obtain the Feynman rules for the three point vertex. As usual, the vertex term from the first term in the Hamiltonian is given by,

$$
-i\lambda
$$

The vertex term from the second term is,

$$
\begin{aligned}
-i\frac{\lambda}{2!}\theta^{\mu\nu}\theta^{\alpha\beta}\left( q_{\mu}q_{\alpha}r_{\nu}r_{\beta} + r_{\mu}r_{\alpha}q_{\nu}q_{\beta} + p_{\mu}p_{\alpha}r_{\nu}r_{\beta} + p_{\mu}p_{\alpha}q_{\nu}q_{\beta} + r_{\nu}r_{\alpha}p_{\nu}p_{\beta} + q_{\mu}q_{\alpha}p_{\nu}p_{\beta}\right)
\end{aligned}
$$

Finally, the vertex contribution of the final term is,

$$
-i\frac{\lambda}{3}\theta^{\mu\nu}\theta^{\alpha\beta}\left(p_{\alpha}p_{\mu}q_{\beta}r_{\nu} + p_{\alpha}p_{\mu}r_{\beta}q_{\nu}+q_{\alpha}q_{\mu}p_{\beta}r_{\nu} + r_{\alpha}r_{\nu}p_{\beta}q_{\nu} + q_{\alpha}q_{\nu}r_{\beta}p_{\nu} + r_{\alpha}r_{\nu}q_{\beta}p_{\nu}\right)
$$

Here, the convention adopted is that $p,q,r$ are all incoming momenta into the vertex. 

For the sake of future brevity, the following tensor may be defined,

\begin{equation}
\begin{aligned}
 Q_{\alpha\beta\mu\nu}&(p,q,r) = \\& \frac{1}{3}\left(p_{\alpha}p_{\mu}q_{\beta}r_{\nu} + p_{\alpha}p_{\mu}r_{\beta}q_{\nu}+q_{\alpha}q_{\mu}p_{\beta}r_{\nu} + r_{\alpha}r_{\nu}p_{\beta}q_{\nu} + q_{\alpha}q_{\nu}r_{\beta}p_{\nu} + r_{\alpha}r_{\nu}q_{\beta}p_{\nu}\right)\\
 & +\frac{1}{2!}\left( q_{\mu}q_{\alpha}r_{\nu}r_{\beta} + r_{\mu}r_{\alpha}q_{\nu}q_{\beta} + p_{\mu}p_{\alpha}r_{\nu}r_{\beta} + p_{\mu}p_{\alpha}q_{\nu}q_{\beta} + r_{\nu}r_{\alpha}p_{\nu}p_{\beta} + q_{\mu}q_{\alpha}p_{\nu}p_{\beta}\right)
\end{aligned}
\end{equation}

Hence, at a vertex, the effective coupling constant is now given by,

\begin{equation}
    -i\lambda\left(g + \theta^{\mu\nu}\theta^{\alpha\beta}Q_{\alpha\beta\mu\nu}(p,q,r)\right)
\end{equation}

\section{Corrections to Scattering}

In this section, we study the corrections to $\phi\phi\rightarrow\phi\phi$ obtained on account of the effective vertex. In this process, labelling the momenta $p_1 p_2 \rightarrow p_3 p_4$, the $s$, $t$ and $u$ channels contribute.

We now note the separate contributions. For the $s$ channel we obtain the term,

\begin{equation}
    i\lambda^{2}\frac{\left(1 + \theta^{\mu\nu}\theta^{\alpha\beta}Q_{\alpha\beta\mu\nu}(p_1, p_2 , -p_1 - p_2)\right)\left(1 + \theta^{\mu\nu}\theta^{\alpha\beta}Q_{\alpha\beta\mu\nu}(-p_3, -p_4 , p_1 + p_2)\right)}{s^2 - m^2}
\end{equation}

For the $t$ channel we have,

\begin{equation}
    i\lambda^{2}\frac{\left(1 + \theta^{\mu\nu}\theta^{\alpha\beta}Q_{\alpha\beta\mu\nu}(p_1, -p_3 , -p_1 + p_3)\right)\left(1 + \theta^{\mu\nu}\theta^{\alpha\beta}Q_{\alpha\beta\mu\nu}(p_2, -p_4 , p_1 - p_3)\right)}{t^2 - m^2}
\end{equation}

For the $u$ channel,

\begin{equation}
    i\lambda^{2}\frac{\left(1 + \theta^{\mu\nu}\theta^{\alpha\beta}Q_{\alpha\beta\mu\nu}(p_1, -p_4 , -p_1 + p_4)\right)\left(1 + \theta^{\mu\nu}\theta^{\alpha\beta}Q_{\alpha\beta\mu\nu}(p_2, -p_3 , p_1 - p_4)\right)}{u^2 - m^2}
\end{equation}

The final amplitude is the squared sum of the above contributions times an overall momentum conserving delta function term. Before proceeding, we note that a certain simplication has been made. In the calculation of the Feynman diagrams, we have assumed that the $\theta^{\mu\nu}$ are $c$ numbers, allowing us to enforce momentum conservation through a delta function obtained by Fourier transforming. If however the $\theta^{\mu\nu}$ are position dependent, further considerations must be made.

As a first step, note that in the limit where $p_1$ is much greater than $p_2$ and $p_4$ is much greater than $p_3$, the $Q_{\alpha\beta\mu\nu}$ reduce to polynomials in $p_1$ and $p_4$, if we keep momentum conservation. The highest such power goes as $\sim p^4$, which diverges for large $p$. Hence, it is not unwarranted to hope that $\theta^{\mu\nu}$ have some position dependence, which upon Fourier transforming, softens this divergence.

The following assumption is now made. If $\theta^{\mu\nu}$ factorises into a position dependent scalar $f$ and a $c$ number part $b^{\mu\nu}$, we only need to study the properties of the Fourier transform of $f$. In the rest of this section, we will study different possibilities of this function and the relevant analytic properties. 

First of all, we note that there will now be a change in the amplitude. To illustrate the change, the $s$ channel diagram contribution becomes,

\begin{equation}
   \int i\lambda^{2}\frac{F^{1}_s F^2_s}{p_5^2 - m^2} \mathrm{d}^4 p_5
\end{equation}

where,

\begin{equation}
    F^1_s = \delta^4(p_1 + p_2 - p_5) + b^{\mu\nu}b^{\alpha\beta}Q_{\alpha\beta\mu\nu}(p_1,p_2,-p_5)\tilde{f}(p_1 + p_2 - p_5)
\end{equation}

\begin{equation}
    F^2_s = \delta^4(p_3 + p_4 - p_5) + b^{\mu\nu}b^{\alpha\beta}Q_{\alpha\beta\mu\nu}(-p_3,-p_4,p_5)\tilde{f}(-p_3 - p_4 + p_5).
\end{equation}

Here, $p_5$ is the momentum communicated by the intermediate scalar particle. Here, we may note that the term of fourth order in $b$ carries with it no delta function. Hence, momentum conservation is not obeyed. Without entering into excessive speculation with regards to possible contributions, we simply note that this corresponds to an inelastic component of the scattering. Before considering a possible expression for the function, we note that in the remaining three terms, delta function momentum conservation holds, and the only contribution factor is $f(0)$.

Now, with regards to the inelastic term, for the sake of concreteness, we consider an example. We may take a gaussian for example,

\begin{equation}
    \tilde{f}(p) = f_{0}(E)f_{1}(p_x)f_{2}(p_y)f_{3}(p_z)
\end{equation}

Each of these is a gaussian of the form,

$$
g(x) = \frac{1}{2\pi^{1/2}\sigma}e^{x^2/\sigma^2}
$$

A couple of cases may be noted. If there is a rapid gain of momentum, such that $p_4$ is much greater than the remaining three, there is a nonzero contribution, weighted by $e^{-E^2/\sigma_E^2 - |p|^2/\sigma^2_{z}}$ in the frame where $p_4$ is along the $z$-axis again showing that this indeed corresponds to an inelastic component. This is heavily damped for highly ultraviolet $p_4$.

Conversely, if one considers a rapid decay of one infrared and one ultraviolet scalar into two infrared scalars, the process is analogous to the one described previously and there is again a small, but nonzero contribution. 

Finally, we may note the case of $\sigma\rightarrow 0$. In this case, four-momentum conservation is again introduced.

Here, we conclude the discussion of noncommutative corrections to scattering in cubic scalar field theory. A number of directions have not been pursued in this article. Firstly, we may consider loop corrections. One may note that a loop correction will give an eighth order contribution to $2\rightarrow 2$ scattering. Hence, at the same order, we have to consider third and fourth order corrections to the vertex at tree level. 

Vertex corrections may also be calculated. These supply sixth order corrections to each vertex. Hence, in calculating eighth order corrections to scattering, we can include vertex corrections where one vertex has a sixth order correction and the other a tree level second order correction.



\end{document}